\documentclass[12pt,a4paper]{article}
\usepackage{graphicx}
\usepackage{subeqnarray}
\begin{document}
%
%                         __________________ 
%                         |                |   
%                         |                |
%                         |  VERSIONE  A   |
%                         |                |
%                         |                |
%                         |                |
%                         |                |
%                         |                |
%                         __________________
%
%                      
%
\newenvironment{lefteqnarray}{\arraycolsep=0pt\begin{eqnarray}}
{\end{eqnarray}\protect\aftergroup\ignorespaces}
\newenvironment{lefteqnarray*}{\arraycolsep=0pt\begin{eqnarray*}}
{\end{eqnarray*}\protect\aftergroup\ignorespaces}
\newenvironment{leftsubeqnarray}{\arraycolsep=0pt\begin{subeqnarray}}
{\end{subeqnarray}\protect\aftergroup\ignorespaces}
\newcommand{\diff}{{\rm\,d}}
\newcommand{\pprime}{{\prime\prime}}
\newcommand{\szeta}{\mskip 3mu /\mskip-10mu \zeta}
\newcommand{\srho}{\mskip 3mu /\mskip-10mu \rho}
\newcommand{\sr}{\mskip 3mu /\mskip-9mu r}
\newcommand{\sR}{\mskip 3mu /\mskip-11mu R}
\newcommand{\sV}{\mskip 3mu /\mskip-10mu V}
\newcommand{\sP}{\mskip 3mu /\mskip-10mu p}
\newcommand{\sPM}{\mskip 3mu /\mskip-11mu P}
\newcommand{\sT}{\mskip 3mu /\mskip-9mu T}
\newcommand{\FC}{\mskip 0mu {\rm F}\mskip-10mu{\rm C}}
\newcommand{\appleq}{\stackrel{<}{\sim}}
\newcommand{\appgeq}{\stackrel{>}{\sim}}
\newcommand{\quadr}{\overline\sqcup}
\newcommand{\Int}{\mathop{\rm Int}\nolimits}
\newcommand{\Nint}{\mathop{\rm Nint}\nolimits}
\newcommand{\arcsinh}{\mathop{\rm arcsinh}\nolimits}
\newcommand{\range}{{\rm -}}
%newcommand{\erf}{\mathop{\rm erf}\nolimits}
%\newcommand{\psfc}{\mathop{\rm psfc}\nolimits}
%\newcommand{\Psf}{\mathop{\rm psf}\nolimits}
\newcommand{\displayfrac}[2]{\frac{\displaystyle #1}{\displaystyle #2}}
%
%\begin{titlepage}
%\setcounter{page}{0}
%\headnote{Astron.~Nachr.~000 (2001) 0, 000--000}
%\makeheadline
%
\title{Updating a simple model of lunar recession}
%{Emden-Chandrasekhar, axisymmetric, rigidly rotating polytropes. VI. %\\
%Self-consistency and continuity}

%% Author and Affilations
\author{
%{D.~Bindoni}\footnote{
%{\it Astronomy Department, Padua Univ., Vicolo Osservatorio 3/2,
%I-35122 Padova, Italy}
%email: daniele.bindoni@unipd.it~~~
%fax: 39-049-8278212}
%, {E.~Milanese}\footnote{
%{\it Physics and Astronomy Department, Padua Univ., Vicolo Osservatorio 3/2,
%I-35122 Padova, Italy}
%email: elena.milanese@studenti.unipd.it~~~
%fax: 39-049-8278212}
 {R.~Caimmi}
\footnote{
{\it Physics and Astronomy Department, Padua University,
Vicolo Osservatorio 3/2, I-35122 Padova, Italy.   Affialiated up to September
30th 2014.  Current status: Studioso Senior.   Current position: in retirement
due to age limits.}
\hspace{50mm}
email: roberto.caimmi@unipd.it~~~
fax: 39-049-8278212}
%, {L.~Secco}\footnote{
%{\it Astronomy Department, Padua Univ., Vicolo Osservatorio 3/2,
%I-35122 Padova, Italy}
%email: luigi.secco@unipd.it~~~
%fax: 39-049-8278212}
\phantom{agga}}

\maketitle

\begin{quotation}
\section*{}
\begin{Large}
\begin{center}

Abstract
\end{center}
\end{Large}
\begin{small}
%\noindent\noindent
A classical model of lunar recession is reviewed and updated, where input
parameters are taken or inferred from earlier astronomical computation of
insolation quantities on Earth back to 0.25 Gyr.   Free parameters are (i)
the transition age, $t_a-t^\dagger$, where mean lunar recession velocity,
$\dot a$, is suddenly lowered, and (ii) the merging age, $t_a-t^\ddagger$,
where Earth-Moon distance (EMD), $a$, drops to zero, which relates to
mean lunar recession velocity just after transition, $\dot a^\dagger$.
Predicted mean EMD and length of day (LOD) slightly overstimate their
counterparts from the above mentioned astronomical computation.   Predicted
LOD, where the effect of atmospheric tides and nontidal processes is
considered and supposed
to be time independent, slightly understimates a linear interpolation from
paleontological data related to Phanerozoic, inferred in earlier
investigations.   If short-period ($\Delta t\approx10^{-3}$ Gyr) EMD
fluctuations ($\Delta a\approx\mp0.5R_\oplus$), resulting from the above
mentioned astronomical computation back to 0.25 Gyr, occur along the whole
evolution of Earth-Moon system (EMS), then the predicted mean EMD
is consistent with values inferred from three well studied
paleontological data sets: Elatina-Reynella (ER), Big Cottonwood (BC), Weeli
Wolli (WW), for $t_a-t^\dagger=0.25$
Gyr and $t_a-t^\ddagger=6$ Gyr.   The same holds for LOD where, in
addition, computed values are consistent with a linear interpolation from
paleontological data related to Proterozoic, inferred in earlier
investigations.   The situation is reversed if the effect of atmospheric tides
and nontidal processes is taken into consideration and suppposed to be time
independent, in the sense that fitting curves relate to merging age in
advance with respect to formation age.   Accordingly, the above
mentioned effect cannot be supposed as time independent during Proterozoic.
In conclusion, the current model can be considered as a useful
zeroth-order approximation to the evaluation of lunar recession and LOD.
More accurate paleontological data, expecially in connection with error
evaluation, would be desirable in view of improved results and further
constraints involving both Astronomy and Paleontology.

%\noindent\noindent
{\it keywords -
Earth-Moon system - tidal friction - lunar recession - length of day.}
\end{small}
\end{quotation}

\section{Introduction} \label{intro}

%\noindent\noindent
The evolution of Earth-Moon system (EMS) has been debated for centuries since
the beginning of the scientific revolution.   Related references are mentioned
in earlier investigations e.g.,
\cite{bro84}\cite{vaa98}\cite{wil00}\cite{laa04} and shall not be repeated
here.

The driving
mechanism lies in the lunar tidal friction, which acts along the following
steps e.g., \cite{wil00}.
\begin{description}
\item[(1)~] 
The gravitational force of the Moon raises a tidal bulge in the solid Earth
and oceans.
\item[(2)\hspace{2mm}]
Because of friction there is a delay in Earth's response, causing the tidal
bulge to lead the Earth-Moon axis by a small angle.
\item[(3)~]
The Moon exerts a torque on the tidal bulge that retards Earth's rotation,
thereby increasing the length of day (LOD).
\item[(4)~]
The torque that Earth's tidal bulge exerts on the Moon leads to an
acceleration of the Moon's orbital motion, causing the Moon to recede from
Earth.
\end{description}
While lunar recession is almost entirely due to tidal friction, LOD can be
affected, in addition, by other processes such as core-mantle friction,
atmospheric tides, mantle convection, macrodiffusion at the core-mantle
boundary e.g., \cite{maj95}\cite{lam80}\cite{dea11}.   Tidal despinning is
less or more effective according if continents are gathered to form a
supercontinent, Pangaea say, or are spread over Earth surface, respectively
e.g., \cite{vaa98}\cite{dea11}.

Direct determination of lunar recession by laser ranging \cite{dia94} yields
a value that mildly changes back to 0.25 Gyr \cite{laa04}.   Extrapolation to
earlier ages would imply catastrophic Earth-Moon close approach within 2 Gyr
or less \cite{lam80}\cite{waz86}\cite{wil00}, which is in contradiction with
inferred age of lunar rocks (brought back via Apollo missions) equal to about
4.5 Gyr e.g., \cite{kop74}\cite{val06}\cite{asp14}.   Then the mean lunar
recession before Phanerozoic (back to about 0.57 Gyr)
had to be significantly lower than the current value and even the average
value through Phanerozoic e.g., \cite{waz86}.

Paleotidal and paleorotational values can be inferred from fossil corals and
molluscs e.g., \cite{scr78}\cite{lam80}\cite{waz86}\cite{soa96b}\cite{sac98}\cite{wil89a}\cite{wil00}.
More specifically, fine laminae are
interpreted either as daily growth increments or as the record of the
semidiurnal or diurnal tide, in accordance with the general growth habits of
related current descendants.   Modulation of the fine banding by the
fortnightly or monthly tidal cycles and by the yearly seasonal cycle allows
estimates of the number of days per year, the number of days per month, and
the number of months per year \cite{waz86}.

In general, paleontological data are presented with no reliable error bands
for both related age and inferred quantities which are quite large anyway
e.g., \cite{vaa98}.   An exception is concerned with at least three cases
concerning precambrian cyclic rhytmites found on the following formations: (i)
Weeli Wolli (WW), Western Australia, about 2.45 Gyr ago
\cite{waz86}\cite{wil89c}\cite{wil90}; (ii) Big Cottonwood (BC), Utah, about
0.90 Gyr ago \cite{soa96b}\cite{sac98}; (iii) Elatina and Reynella Siltstone
(ER), South Australia, about 0.62 Gyr ago
\cite{wil89a}\cite{wil89b}\cite{wil89c}\cite{wil90}\cite{wil91}\cite{wil94}\cite{wil97}.
For further details, an
interested reader is addressed to the parent paper \cite{wil00}.

A nontrivial question is how to consider the above mentioned paleontological
data with respect to the remaining ones.   To this respect, an extreme case
could be excluding the latter e.g., \cite{wil00} and the opposite could be
assigning equal statistical weight to all available data e.g., \cite{dea11}.
Comparison of model predictions to values inferred from observations would
imply the former alternative, in that error bars are needed, while
determination of empirical laws would prefer the latter alternative, where
larger datasets are available.

Under restrictive but plausible assumptions, the mean Earth-Moon distance
(EMD), $a$, in the past can be expressed in terms of the age, $t_a-t$
($t_a$ is the current EMS age), the current EMD, $a_a$,
and the current mean lunar recession, $<\dot a_a>$, which implies catastrophic
Earth-Moon collision within 2 Gyr or less
\cite{lam80}\cite{waz86}\cite{wil00}.  To avoid
this discrepancy, a sudden reduction in mean lunar recession velocity at an
assigned transition age is necessary \cite{waz86}\cite{wil00}.
Accordingly, model key parameters
are the transition age, $t_a-t^\dagger$, and the merging
age, $t_a-t^\ddagger$, where the latter has necessarily to take place in
advance with respect to EMS formation age, $t_a-t_i$.

In this view, the merging age is an artefact of the model, in the sense the
Moon orbit becomes unstable when Roche limit or corotation limit is attained,
and computation should be halted there.   In addition, EMS birth has to be
conceived as the earliest configuration where (quasi) equilibrium holds.   To
this respect, all proposed models of Moon formation e.g., \cite{asp14} could
be viable.

Though the above mentioned classical model of lunar recession has
been used in the past e.g., \cite{waz86}\cite{wil00}, still detailed
formulation and exploration of parameter space is lacking (to the knowledge of
the author) and it could be a useful zeroth order approximation for more
refined attempts.   In addition, Moon formation is never explicitly mentioned
(to the knowledge
of the author) in dealing with paleotidal and paleorotational values which, on
the other hand, could strenghten mutual connection between astronomical and
paleontological data.

The current paper is aimed to review and update the classical model of lunar
recession e.g., \cite{waz86}\cite{wil00}, according to the above
considerations.    The model and related input parameters are presented in
Section \ref{mod} after a number of basic considerations.   The results are
shown in Section \ref{resu} and compared
with their counterparts inferred from both astronomical computations (0-0.25
Gyr ago) \cite{laa04} and paleontological data (0-2.5 Gyr ago)
\cite{waz86}\cite{vaa98}\cite{wil00}\cite{dea11}.
The discussion is performed in
Section \ref{disc} and the conclusion makes the subject of Section \ref{conc}.

\section{A classical model of lunar recession}
\label{mod}
\subsection{Boundary conditions}
\label{boco}

%\noindent\noindent
Let a two-body system be considered where the secondary, S (in particular, a
satellite) undergoes circular orbits around the primary, P (in particular, a
planet).   Let the two bodies be rigid and spherical-symmetric.  If $a_1$ and
$a_2$ are radii of different orbits, the application of Kepler third law
yields:
\begin{lefteqnarray}
\label{eq:K3}
&& \left(\frac{P_1}{P_2}\right)^2=\left(\frac{a_1}{a_2}\right)^3~~;
\end{lefteqnarray}
where $P$ is the orbital period.   In presence of satellite (secular)
recession, the indexes, 1 and 2, can be related to different ages e.g.,
\cite{deu90}.

In the special case of negligible (with respect to unity) secondary-to-primary
mass ratio, the lower limit of stable orbits, or Roche limit, is e.g.,
\cite{jea29} Chap.VIII \S212:
\begin{lefteqnarray}
\label{eq:Rl}
&& a_{\rm R}=2.4554\left(\frac{\overline\rho_{\rm P}}{\overline\rho_{\rm S}}
\right)^{1/3}R_{\rm P}~~;\qquad\frac{M_{\rm S}}{M_{\rm P}}\ll1~~;
\end{lefteqnarray}
where $a_{\rm R}$ is the orbital radius when the Roche limit is attained,
$\overline\rho$ is the mean density, $R$ the radius and $M$ the mass.   After
performing little algebra, Eq.\,(\ref{eq:Rl}) translates into:
\begin{lefteqnarray}
\label{eq:RlM}
&& a_{\rm R}=2.4554\left(\frac{M_{\rm P}}{M_{\rm S}}\right)^
{1/3}R_{\rm S}~~;\qquad\frac{M_{\rm S}}{M_{\rm P}}\ll1~~;
\end{lefteqnarray}
in terms of masses instead of mean densities.   The numerical coefficient on
the right-hand side of Eqs.\,(\ref{eq:Rl}) and (\ref{eq:RlM}) can be
approximeted as $2^{4/3}\approx2.5198$.

The orbital period of a stable ($a_{\rm S}\ge a_{\rm R}$) orbit is e.g.,
\cite{asp14}:
\begin{lefteqnarray}
\label{eq:PS}
&& P_{\rm S}=2\pi\left(\frac{a_{\rm S}^3}{GM_{\rm P}}\right)^{1/2}~~;\qquad
M_{\rm S}\ll M_{\rm P}~~;
\end{lefteqnarray}
where $G=6.67408\cdot10^{-8}$ cm$^3$/(g\,s$^2$) is the constant of
gravitation.
In the Roche limit ($a_{\rm S}=a_{\rm R}$) Eq.\,(\ref{eq:PS}) reads:
\begin{lefteqnarray}
\label{eq:PR}
&& P_{\rm R}=2\pi\left(\frac{a_{\rm R}^3}{a_{\rm S}^3}
\frac{a_{\rm S}^3}{GM_{\rm P}}\right)^{1/2}=
\left(\frac{a_{\rm R}}{a_{\rm S}}\right)^{3/2}P_{\rm S}~~;\qquad
M_{\rm S}\ll M_{\rm P}~~;
\end{lefteqnarray}
in agreement with Eq.\,(\ref{eq:K3}).

With regard to EMS, $R_{\rm S}=R_{\bigcirc}=1378$\,km,
$M_{\rm S}/M_{\rm P}=M_\bigcirc/M_\oplus=0.0123\ll1$,
and Eq.\,(\ref{eq:RlM}) reduces to:
\begin{lefteqnarray}
\label{eq:aRl}
&& a_{\rm R}=18487.1467\,{\rm km}=2.8985\,R_\oplus~~;
\end{lefteqnarray}
where $R_{\bigcirc}$, $M_{\bigcirc}$, $R_{\rm P}=R_\oplus$, are Moon radius,
Moon mass, Earth radius, respectively.

The particularization of Eq.\,(\ref{eq:K3}) to the current EMS configuration,
via Eq.\,(\ref{eq:aRl}) yields:
\begin{lefteqnarray}
\label{eq:PRl}
&& P_{\rm R}=6.9487\,{\rm h}~~;
\end{lefteqnarray}
where $a_\bigcirc=60.1426 R_\oplus$ and $P_\bigcirc=27.4404$\,d have been
inferred from an earlier investigation \cite{laa04}.   If EMS evolution is
intended as a sequence of (quasi) equilibrium configurations, then lunar
distance and orbital period are confined above a threshold of about
$3R_\oplus$ and 7\,h, respectively.

Turning to the general case, let $T_{\rm P}$ be the rotation period of the
primary body.   The corotation condition is $P_{\rm S}=T_{\rm P}$ e.g.,
\cite{asp14} which, via Eq.\,(\ref{eq:PS}), after little algebra reads:
\begin{lefteqnarray}
\label{eq:aC}
&& a_{\rm C}=\left(\frac{GM_{\rm P}}{4\pi^2}\right)^{1/3}T_{\rm P}^{2/3}~~;
\end{lefteqnarray}
where $a_{\rm S}=a_{\rm C}$ is the orbital radius in presence of corotation.
If the Roche limit coincides with the corotation limit, $a_{\rm R}=a_{\rm C}$,
then $T_{\rm P}=P_{\rm R}$ via Eqs.\,(\ref{eq:PR}) and (\ref{eq:aC}).   A
secondary body orbitating below the corotation radius will spiral in towards
the primary due to tides e.g., \cite{asp14}.

With regard to EMS, $M_{\rm P}=M_\oplus=5.9726\cdot10^{27}$\,g and
Eq.\,(\ref{eq:aC}) reduces to:
\begin{lefteqnarray}
\label{eq:aCM}
&& a_{\rm C}=0.795989R_\oplus\left(\frac T{\rm h}\right)^{2/3}~~;
\end{lefteqnarray}
where $T=T_\oplus$ is LOD at an assigned age.

For fluid masses in rigid rotation, a necessary condition for equilibrium
(due to Poincar\'e) is e.g., \cite{jea29} Chap.\,IX \S240:
\begin{lefteqnarray}
\label{eq:OmP}
&& \Omega^2<\Omega_{\rm P}^2=2\pi G\overline\rho_{\rm P}~~;
\end{lefteqnarray}
where $\Omega$ is the angular velocity and the index, P, denotes Poincar\'e
limit.   Related Poincar\'e rotation period,
$T_{\rm P}=2\pi/\Omega_{\rm P}$, is inferred from Eq.\,(\ref{eq:OmP}) as:
\begin{lefteqnarray}
\label{eq:TPo}
&& T_{\rm P}=\frac{2\pi}{(2\pi G\overline\rho)^{1/2}}=2\pi\left(\frac23\frac
{R^3}{GM}\right)^{1/2}~~;
\end{lefteqnarray}
where the last equality holds for spherical bodies.   Shorter periods,
$T<T_{\rm P}$, would imply fission or equatorial rings in centrifugal support
e.g., \cite{jea29} Chap.\,IX \S\S235-240.   With regard to the Earth,
$R=R_\oplus=6378.1366\cdot10^5$ cm and Eq.\,(\ref{eq:TPo}) reduces to:
\begin{lefteqnarray}
\label{eq:TPE}
&& T_{\rm P}=1.149728\,{\rm h}~~;
\end{lefteqnarray}
which is a LOD lower limit for Earth stability.

In summary, sequences of EMS equilibrium configurations must necessarily
satisfy the following boundary conditions:
\begin{lefteqnarray}
\label{eq:bca}
&& a\ge a_{\rm R}=2.898518 R_\oplus~~; \\
\label{eq:bcP}
&& P\ge P_{\rm R}=6.948719\,{\rm h}~~;
\end{lefteqnarray}
radius and period of lunar orbit above Roche limit;
\begin{lefteqnarray}
\label{eq:bcC}
&& a\ge a_{\rm C}=0.795989 R_\oplus~~;
\end{lefteqnarray}
radius of lunar orbit above corotation radius;
\begin{lefteqnarray}
\label{eq:bcT}
&& T\ge T_{\rm P}=1.149728\,{\rm h}~~;
\end{lefteqnarray}
LOD above Poincar\'e limit.

\subsection {The model}
\label{cmlr}

According to the classical model of lunar recession e.g.,
\cite{lam80}\cite{waz86}\cite{deu90}\cite{vaa98}\cite{wil00}\cite{dea11}, the
following restrictions hold.
\begin{description}
\item[(1)~\,] 
EMS evolution occurs along a sequence of (quasi) equilibrium configurations
where the total angular momentum remains unchanged.
\item[(2)~] 
The driving mechanism is due to semidiurnal tidal friction on solid Earth +
ocean, which causes angular momentum transfer from Earth spin to Moon
revolution, increasing both LOD and lunar recession.
\item[(3)~]
The tidal friction from the Sun is negligible with respect to the tidal
friction from the Moon.
\item[(4)~\,]
Earth and Moon mass distribution are spherical-symmetric in absence of tidal
effects.
\item[(5)~\,] 
Earth and Moon mass and radius remain unchanged.
\item[(6)~\,] 
The Moon retains syncronous rotation.
\item[(7)~~] 
Moon orbit retains circular and lying on the ecliptic plane.
\end{description}
For further details, an interested reader is addressed to the above quoted
papers.

Within the framework of the classical model, EMD at the age, $t_a-t$,
e.g., \cite{waz86}\cite{wil00} and related recession velocity read:
\begin{lefteqnarray}
\label{eq:a}
&& \frac a{a_a}=\left[1-\frac{13}2(t_a-t)\frac{<\dot a_a>}{a_a}\right]^{2/13}
~~; \\
\label{eq:ad}
&& \frac{\dot a}{<\dot a_a>}=\left[1-\frac{13}2(t_a-t)\frac{<\dot a_a>}{a_a}
\right]^{-11/13}~~;
\end{lefteqnarray}
and the inverse function is:
\begin{lefteqnarray}
\label{eq:t}
&& t_a-t=\frac2{13}\frac{a_a}{<\dot a_a>}\left[1-\left(\frac a{a_a}\right)^
{13/2}\right]~~;
\end{lefteqnarray}
where extrapolation to merging configuration, $a=0$, yields:
\begin{lefteqnarray}
\label{eq:t0}
&& t_a-t^\ddagger=\frac2{13}\frac{a_a}{<\dot a_a>}~~;
\end{lefteqnarray}
which marks the end of the sequence.

The substitution of Eq.\,(\ref{eq:t0}) into (\ref{eq:a}) after little algebra
yields:
\begin{lefteqnarray}
\label{eq:aag}
&& \frac a{a_a}=\left[1-\frac{t_a-t}{t_a-t^\ddagger}\right]^
{2/13}=\left[\frac{(t_a-t^\ddagger)-(t_a-t)}{t_a-t^\ddagger}
\right]^{2/13}~~;
\end{lefteqnarray}
in terms of the merging age, $t_a-t^\ddagger$.

Merging configurations must be conceived as purely fictitious (unless EMS took
birth from fission), keeping in mind sequences of (quasi) equilibrium
configurations necessarily end by attaining the Roche limit.   On the other
hand, the merging age cannot occur in advance of EMS formation, as:
\begin{lefteqnarray}
\label{eq:tfB}
&& t_a-t_i\le t_a-t^\ddagger~~;
\end{lefteqnarray}
and sequences for which Eq.\,(\ref{eq:tfB}) is violated must be disregarded.

Using current values of EMD and mean lunar recession velocity, listed in Table
\ref{t:inpu}, the merging age via Eq.\,(\ref{eq:t0}) reads:
\begin{lefteqnarray}
\label{eq:t0M}
&& t_a-t^\ddagger=1.516617\,{\rm Gyr}~~;
\end{lefteqnarray}
which is at odds with Eq.\,(\ref{eq:tfB}), keeping in mind
$t_a-t_i\approx4.5$ Gyr e.g., \cite{asp14}.
The above discrepancy could be avoided, provided mean lunar recession velocity
was subsantially reduced in the past with respect to the present e.g.,
\cite{waz86}\cite{wil00}.

More specifically, the simplest option is a single discontinuity at a \
transition age, $t_a-t^\dagger$, where $\dot a(t_a-t^\dagger)$ instantaneously
drops to $\dot a^\dagger$.   Accordingly, Eq.\,(\ref{eq:a}) is valid back to
$t_a-t^\dagger$, as:
\begin{lefteqnarray}
\label{eq:adg}
&& \frac{a^\dagger}{a_a}=\left[1-\frac{13}2(t_a-t^\dagger)\frac{<\dot a_a>}
{a_a}\right]^{2/13}~~;
\end{lefteqnarray}
in earlier ages, $t_a-t>t_a-t^\dagger$, Eqs.\,(\ref{eq:a}), (\ref{eq:t0}),
(\ref{eq:aag}), translate into:
\begin{lefteqnarray}
\label{eq:aad}
&& \frac a{a^\dagger}=\left[1-\frac{13}2(t^\dagger-t)\frac{<\dot a^\dagger>}
{a^\dagger}\right]^{2/13}~~; \\
\label{eq:tdg}
&& t^\dagger-t^\ddagger=\frac2{13}\frac{a^\dagger}{<\dot a^\dagger>}~~; \\
\label{eq:atg}
&& \frac a{a^\dagger}=\left[1-\frac{t^\dagger-t}{t^\dagger-t^\ddagger}\right]^
{2/13}=\left[\frac{(t^\dagger-t^\ddagger)-(t^\dagger-t)}{t^\dagger-t^\ddagger}
\right]^{2/13}~~;
\end{lefteqnarray}
where $t^\dagger-t$ is the age since the transition age, $t_a-t^\dagger$.
Accordingly, free parameters of the model are the transition age and the
merging age.

Under the restrictive assumption LOD changes only via tidal friction, using
angular momentum conservation and Kepler's third law yields \cite{deu90}%
\footnote{The term, $(a_0/a)^6$ has to be replaced by $(a/a_0)^6$ in Eq.\,(2)
therein, to be consistent with next Eq.\,(3).}:%
\begin{lefteqnarray}
\label{eq:aO}
&& \left(\frac a{a_a}\right)^{1/2}+\frac{C_2}{13}\left(\frac a{a_a}\right)^
{13/2}=-C_1\frac\Omega{\Omega_a}+C_3~~; \\
\label{eq:c123}
&& C_1=\frac1{4.93}\,;~C_2=(0.46)^2=0.2116\,;~C_3=1+C_1+\frac{C_2}{13}=
1.219117\,;
\end{lefteqnarray}
where $\Omega$ is Earth angular velocity, $C_1$ is the current ratio of Earth
spin angular momentum to Moon orbital angular momentum, $C_2$ is the current
ratio of solar to lunar retarding torques acting on Earth, and $C_3$ is an
integration constant e.g., \cite{deu90}.

In terms of LOD, $T=2\pi/\Omega$, Eq.\,(\ref{eq:aO}) after some algebra reads:
\begin{lefteqnarray}
\label{eq:T}
&& \frac T{T_a}=C_1\left\{C_3-\left(\frac a{a_a}\right)^{1/2}\left[1+\frac
{C_2}{13}\left(\frac a{a_a}\right)^6\right]\right\}^{-1}~~;
\end{lefteqnarray}
and LOD variation rate after some algebra can be inferred from
Eq.\,(\ref{eq:T}) as:
\begin{lefteqnarray}
\label{eq:Td}
&& \frac{\dot T}{T_a}=\frac12\frac1{C_1}\frac{T^2}{T_a^2}\left(\frac a{a_a}
\right)^{-1/2}\frac{\dot a}{a_a}\left[1+C_2\left(\frac a{a_a}\right)^6\right]
~~;
\end{lefteqnarray}
where the current value is:
\begin{lefteqnarray}
\label{eq:Tda}
&& \frac{\dot T_a}{T_a}=\frac12\frac1{C_1}\frac{\dot a_a}{a_a}(1+C_2)~~;
\end{lefteqnarray}
finally, dividing each side of Eq.\,(\ref{eq:Td}) by its counterpart of
Eq.\,(\ref{eq:Tda}) yields:
\begin{lefteqnarray}
\label{eq:Tdda}
&& \frac{\dot T}{\dot T_a}\left(\frac T{T_a}\right)^{-2}=\frac{\dot a}
{\dot a_a}\left(\frac a{a_a}\right)^{-1/2}\frac{1+C_2(a/a_a)^6}{1+C_2}~~;
\end{lefteqnarray}
where $\dot a_a=<\dot a_a>$ for sake of brevity.

With regard to the transition age, $t_a-t^\dagger$, Eqs.\,(\ref{eq:T}) and
(\ref{eq:Tdda}) after some algebra translate into:
\begin{lefteqnarray}
\label{eq:Tp}
&& \frac T{T^\dagger}=C_1^\prime\left\{C_3^\prime-\left(\frac a{a^\dagger}
\right)^{1/2}\left[1+\frac{C_2^\prime}{13}\left(\frac a{a^\dagger}\right)^6
\right]\right\}^{-1}~~; \\
\label{eq:Tpda}
&& \frac{\dot T}{\dot T^\dagger}\left(\frac T{T^\dagger}\right)^{-2}=\frac
{\dot a}{\dot a^\dagger}\left(\frac a{a^\dagger}\right)^{-1/2}\frac{1+C_2
(a/a^\dagger)^6}{1+C_2}~~; \\
\label{eq:Cp123}
&& C_1^\prime=C_1\frac{T_a}{T^\dagger}\left(\frac{a^\dagger}{a_a}\right)^{1/2}
\,;~C_2^\prime=C_2\left(\frac{a^\dagger}{a_a}\right)^6\,;~C_3^\prime=C_3\left(
\frac{a^\dagger}{a_a}\right)^{-1/2}\,;
\end{lefteqnarray}
where Eqs.\,(\ref{eq:Tp}) and (\ref{eq:Tpda}) retain the formal expression of
Eqs.\,(\ref{eq:T}) and (\ref{eq:Tdda}), respectively, regardless of the
transition age, provided related constants are redefined via
Eq.\,(\ref{eq:Cp123}).

In the special case of merging age, $a=0$, Eq.\,(\ref{eq:T}) reduces to:
\begin{lefteqnarray}
\label{eq:T0a}
&& \frac{T^\ddagger}{T_a}=\frac{C_1}{C_3}=0.1663826~~;
\end{lefteqnarray}
accordingly, LOD (under the influence of tidal friction) attains a minimum
value as:
\begin{lefteqnarray}
\label{eq:T0}
&& T\ge T^\ddagger=0.1663826\,T_a=3.982278\,{\rm h}~~;
\end{lefteqnarray}
regardless of the sequence under consideration.

\subsection{Input parameters}
\label{inpa}

For dealing with a self-consistent set of input parameters, values are taken
or inferred
from an earlier investigation \cite{laa04} as listed in Table \ref{t:inpu}.
\begin{table*}
\caption[par]{Values of input parameters used for computations, taken or
inferred
from an earlier investigation with regard to J2000.0 date \cite{laa04}.   The
current lunar period is determined via Eq.\,(\ref{eq:PS}).   The difference
with respect to current values is acceptably low except for $\dot T_a$.}
%\vspace{0.3cm}
\label{t:inpu}
\begin{center}
%\footnotesize
\begin{tabular}{lcl}
\hline
\multicolumn{1}{c}{parameter} &
\multicolumn{1}{c}{value} &
\multicolumn{1}{c}{caption} \\
\hline\noalign{\smallskip}
%                      &              &                                  \\
$R_\oplus$/km         & 6378.1366    & equatorial Earth radius          \\
$R_\bigcirc$/km       & 1738         & equatorial Moon radius           \\
$M_\bigcirc/M_\oplus$ & 0.0123000383 & Moon-to-Earth mass ratio         \\
$T_a$/h               & 23.934468    & current length of day            \\
$\dot T_a$/(ms/cy)    & 2.675580     & current change in length of day  \\
$a_a/R_\oplus$        & 60.142611    & current Earth-Moon distance      \\
$<\dot a_a>$/(cm/y)   & 3.891229     & current lunar recession velocity \\
$P_\bigcirc$/d        & 27.440392    & current lunar orbital period     \\
\noalign{\smallskip}
\hline
\end{tabular}
\end{center}
\end{table*}
EMS evolution therein is considered within the framework of a new solution for
astronomical computation of insolation quantities on Earth spanning from
$-$0.25 Gyr to +0.25 Gyr.   Related polynomial approximations for mean EMD
and LOD read \cite{laa04}:
\begin{leftsubeqnarray}
\slabel{eq:apoa}
&& \frac a{R_\oplus}=\sum_{k=0}^4A_k\left(\frac{t_a-t}{\rm Gyr}\right)^k~~; \\
\slabel{eq:apob}
&& A_0=\frac{a_a}{R_\oplus}=60.142611~;\quad A_1=-6.100887~;\quad
A_2=-2.709407~; \\
\slabel{eq:apoc}
&& A_3=-1.366779~;\quad A_4=-1.484062~;
\label{seq:apo}
\end{leftsubeqnarray}
\begin{leftsubeqnarray}
\slabel{eq:Tpoa}
&& \frac T{\rm h}=\sum_{k=0}^4B_k\left(\frac{t_a-t}{\rm Gyr}\right)^k~~; \\
\slabel{eq:Tpob}
&& B_0=\frac{T_a}{\rm h}=23.934468~;\quad B_1=-7.432167~;\quad
B_2=-0.727046~; \\
\slabel{eq:Tpoc}
&& B_3=-0.409572~;\quad B_4=-0.589692~;
\label{seq:Tpo}
\end{leftsubeqnarray}
where $a$ appears in place of $<a>$ for sake of brevity and the terms related
to $k>4$ are negligible in both cases.

In particular, current mean lunar velocity recession and LOD variation rate
are inferred
from Eqs.\,(\ref{seq:apo}) and (\ref{seq:Tpo}), respectively, as listed in
Table \ref{t:inpu}.   While the former result is consistent with the current
mean lunar velocity recession inferred from lunar laser ranging,
$<\dot a_a>=(3.82\mp0.07$)\,cm/y \cite{dia94}, the
latter substantially exceeds LOD variation rates mentioned in different
investigations e.g., $\dot T_a=(2.3\mp0.1)$\,ms/cy \cite{vaa98}\cite{dea11};
$\dot T_a=(2.50\mp0.01)$\,ms/cy \cite{mal09}.

A comparison between numerical integration and analytical approximation
discloses the occurrence of short-period ($\Delta t\approx10^{-3}$\,Gyr)
EMD fluctuations ($\Delta a/R_\oplus\approx\mp0.5$) around the local mean
value
\cite{laa04}.   Accordingly, EMD evolution lies within a band, centered on
the local mean value, where the vertical width equals about one Earth radius.
A choice of input parameters as listed in Table \ref{t:inpu} leaves two
free parameters, namely the transition age, $t_a-t^\dagger$, and the merging
age, $t_a-t^\ddagger$.

Computer runs were performed for the following cases: (a)
$(t_a-t^\dagger)/{\rm Gyr}=0.57$; $(t_a-t^\ddagger)/{\rm Gyr}=1.516617$ (no
discontinuity in $<\dot a>$), 2.0, 2.5, ..., 4.5, and (b)
$(t_a-t^\dagger)/{\rm Gyr}=0.25$;
$(t_a-t^\ddagger)/{\rm Gyr}=1.516617$ (no discontinuity in $<\dot a>$),
2.0, 2.5, ..., 4.5, 5.0, 6.0, 7.0, 10.0.   More specifically, a discontinuity
in lunar recession velocity takes place (more or less) at the transition
between Proterozoic and Phanerozoic in the former alternative, and at
formation and deformation of Pangaea supercontinent in the latter alternative
e.g., \cite{vaa98}\cite{dea11}.

\section{Results}
\label{resu}

\begin{figure*}[t]               
\begin{center}                   
\includegraphics[scale=0.8]{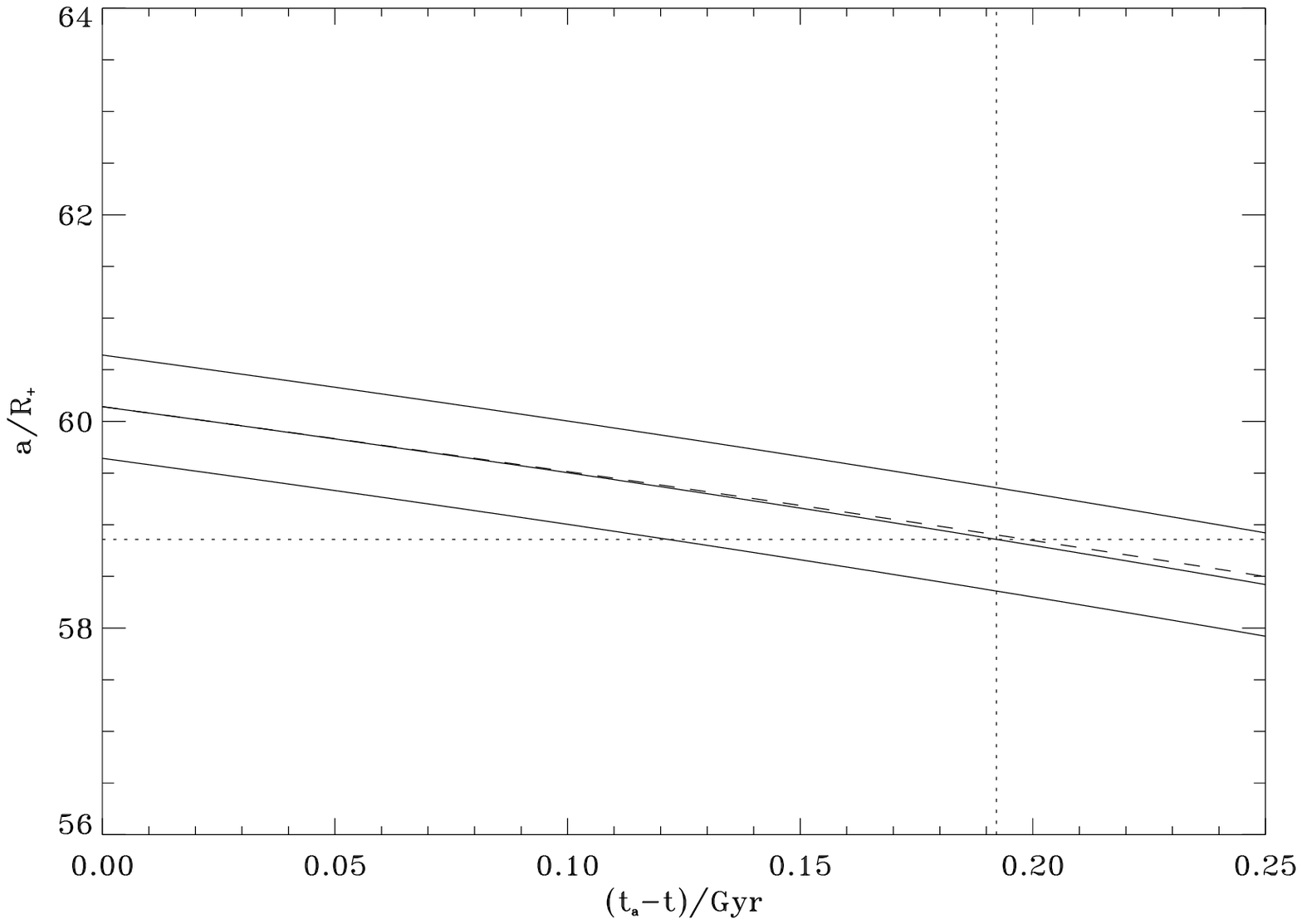}                      
\caption[ddbb]{Predicted mean lunar distance (EMD, in Earth radii),
$a/R_\oplus$,
vs age (Gyr, back in time since J2000.0 date), $t_a-t$, for past 0.25 Gyr,
according to Eq.\,(\ref{eq:a}), dashed line, and Eq.\,(\ref{seq:apo}), full
line within the band.   The band vertical width ($\Delta a/R_\oplus\approx1$)
is owing to short-period ($\Delta t\approx10^{-3}$\,Gyr) EMD fluctuations.
Dotted straight lines define a special point,
$(0.192150014, 58.85857007)$, on the central full curve for which
the ratio of mean lunar distance to lunar diameter reads
$a/(2R_\bigcirc)=108$.   See text for further details.}
\label{f:t108}     
\end{center}       
\end{figure*}

With regard to the past 0.25 Gyr, $0\le(t_a-t)/{\rm Gyr}\le0.25$, 
predicted mean EMD (in Earth radii), $a/R_\oplus$, vs age,
$t_a-t$, according to Eq.\,(\ref{eq:a}), is shown in Fig.\,\ref{f:t108} as a
dashed line e.g., \cite{waz86}\cite{wil00}.
By comparison, mean EMD inferred from astronomical computation within the
framework of a new solution for the insolation quantities on Earth, according
to Eq.\,(\ref{seq:apo}), is shown in Fig.\,\ref{f:t108} as a full line inside
a band \cite{laa04}.   The band vertical width $(\Delta a/R_\oplus\approx1)$
is owing to short-period ($\Delta t\approx10^{-3}$\,Gyr) EMD fluctuations
\cite{laa04}.   An inspection of Fig.\,\ref{f:t108} discloses
Eq.\,(\ref{eq:a}) is in close agreement with its counterpart,
Eq.\,(\ref{seq:apo}), related to a more sophisticated model \cite{laa04}.

Dotted straight lines define a special point, $(0.192150014, 58.85857007)$, on
the central full curve, for which
the ratio of mean EMD to lunar diameter reads $a/(2R_\bigcirc)=108$.   The
occurrence of short-period fluctuations implies the above value can be
attained within a range, $0.12\appleq(t_a-t)/{\rm Gyr}\appleq0.25$.

\begin{figure*}[t]               
\begin{center}                   
\includegraphics[scale=0.8]{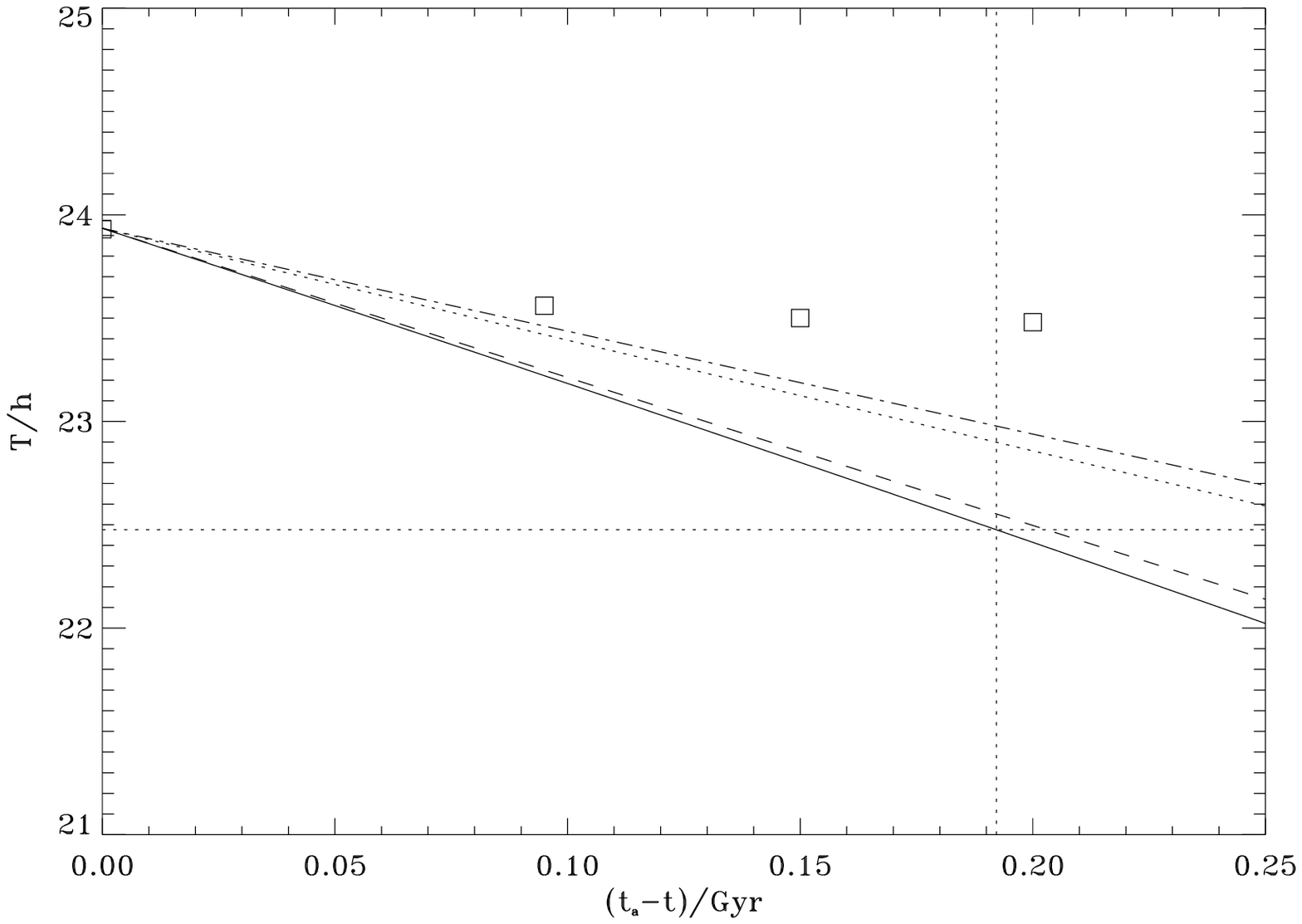}                      
\caption[ddbb]{Predicted length of day (LOD) due to tidal friction, $T/$h,
vs age (Gyr, back in time since J2000.0 date), $t_a-t$, for past 0.25 Gyr,
according to Eq.\,(\ref{eq:T}), dashed line, and Eq.\,(\ref{seq:Tpo}), full
line.   Dotted straight lines define a special point,
$(0.192150014, 22.47646196)$, on the full curve for which
the ratio of mean lunar distance to lunar diameter reads
$a/(2R_\bigcirc)=108$.   A linear interpolation from paleontological data is
plotted as a dot-dashed line.   A nonlinear trend also inferred from
paleontological data is shown as squares.   Predicted LOD including time
independent effects from atmospheric tides and nontidal processes is
represented as a dotted curve.   See text for further details.}
\label{f:tt108}     
\end{center}       
\end{figure*}

Predicted LOD due to tidal friction, $T/$h, vs age, $(t_a-t)/$Gyr, according
to Eq.\,(\ref{eq:T}), is shown in Fig.\,\ref{f:tt108} as a dashed line.
By comparison, LOD inferred from astronomical computation within the
framework of a new solution for the insolation quantities on Earth, according
to Eq.\,(\ref{seq:Tpo}), is shown in Fig.\,\ref{f:tt108} as a full line
\cite{laa04}.   An inspection of Fig.\,\ref{f:tt108} discloses
Eq.\,(\ref{eq:T}) is in close agreement with its counterpart,
Eq.\,(\ref{seq:Tpo}), related to a more sophisticated model \cite{laa04}.

Dotted straight lines define a special point, $(0.192150014, 22.47646196)$, on
the full curve, for which
the ratio of mean EMD to lunar diameter reads $a/(2R_\bigcirc)=108$.   

The dot-dashed line is a linear interpolation from paleontological data,
$T/$h$=T_a/$h$-4.98(t_a-t)/$Gyr, shown in earlier investigations
\cite{vaa98}\cite{dea11}.   A nonlinear trend also inferred from
paleontological data \cite{vaa98}\cite{dea11} is shown as squares.

\begin{figure*}[t]               
\begin{center}                   
\includegraphics[scale=0.8]{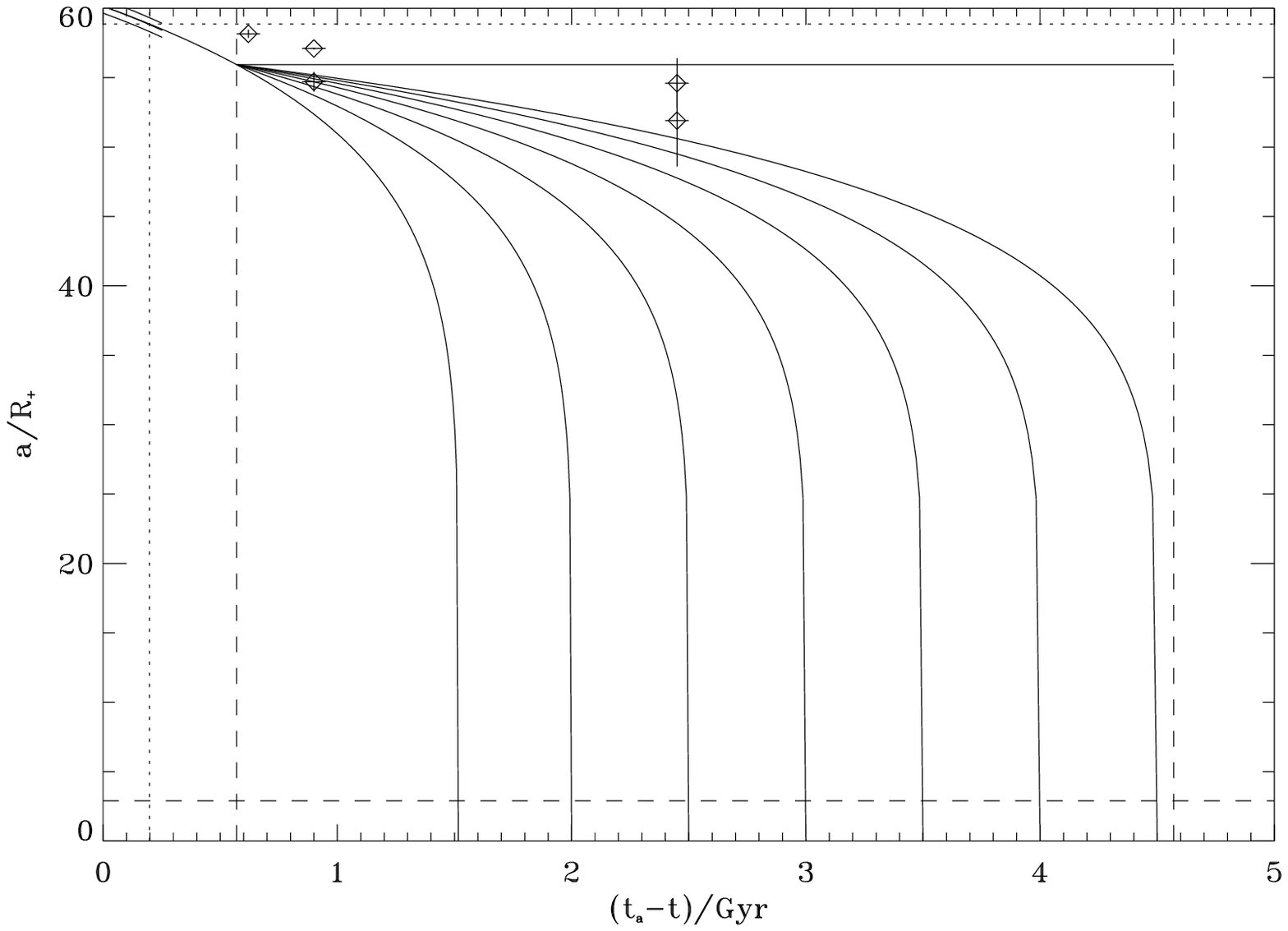}                      
\caption[ddbb]{Predicted mean lunar distance (EMD, in Earth radii),
$a/R_\oplus$,
vs age (Gyr, back in time since J2000.0 date), $t_a-t$, for past 4.57 Gyr,
according to Eq.\,(\ref{eq:a}), for different merging ages (from the left to
the right), $(t_a-t^\ddagger)/$Gyr$=1.516617$, 2.0, 2.5, 3.0, 3.5, 4.0, 4.5,
$+\infty$ (no lunar recession).   All features appearing in Fig.\,\ref{f:t108}
are included within $0\le(t_a-t)/$Gyr$\,\le0.25$, but dotted straight lines
define a special point, (0.198499271, 58.85857007), on the dashed curve
therein, for which the ratio of mean lunar distance to lunar diameter reads
$a/(2R_\bigcirc)=108$.   Dashed vertical straight lines mark the selected
transition age, $(t_a-t^\dagger)/$Gyr$=0.57$ (left) and the assumed age of the
Earth-Moon system (EMS), $(t_a-t_i)/$Gyr$=4.57$ (right).   The Roche limit,
$a_{\rm R}/R_\oplus=2.898518$, is defined by a horizontal straight line.
Values inferred from
accurately studied paleontological data are plotted as diamonds, related to
different authors when appropriate \cite{wil00}.   Source of data (from the
left to the right): ER, BC (lower value preferred), WW (upper value preferred)
\cite{wil00}.   See text (Introduction) for further details.}
\label{f:u108}     
\end{center}       
\end{figure*}

Concerning whole EMS evolution, assumed to start from $(t_a-t_i)/$Gyr$=4.57$,
and a transition age, $(t_a-t^\dagger)/$Gyr$=0.57$, predicted EMD (in Earth
radii), $a/R_\oplus$, vs age, $(t_a-t)/$Gyr, is shown in Fig.\,\ref{f:u108}
as full lines
for selected merging ages, $(t_a-t^\ddagger)$, or selected mean lunar
recession velocities just after the transition age, $<\dot a^\dagger>$, via
Eq.\,(\ref{eq:t0}), as listed in Table \ref{t:dapt}.
\begin{table*}
\caption[par]{Mean lunar velocity recession just after the transition age,
$<\dot a^\dagger>$, for different selected merging ages, $t_a-t^\ddagger$,
with regard to transition ages, $t_a-t^\dagger=0.57$\,Gyr (left column) and
0.25 Gyr (right column).   The first row relates to absence of discontinuity
at the transition age in connection with a merging age,
$t_a-t^\ddagger=1.516617$\,Gyr.   Accordingly, $<\dot a>=5.7982$\,cm/y
(left column) and 4.5319\,cm/y) (right column) just before transition,
regardless of the merging age.   All cases are plotted in
Figs.\,\ref{f:u108}-\ref{f:ut08} (left column) and \ref{f:v108}-\ref{f:vt08}
(right column).}
%\vspace{0.3cm}
\label{t:dapt}
\begin{center}
%\footnotesize
\begin{tabular}{rll}
\hline
\multicolumn{1}{c}{$\frac{t_a-t^\ddagger}{\rm Gyr}$} &
%\multicolumn{1}{r}{$R_{\rm int}$} &
%\multicolumn{1}{r}{$R_{\rm ext}$} &
%\multicolumn{1}{r}{$\Delta R$} &
%\multicolumn{1}{c}{name} &
\multicolumn{2}{c}{$\frac{<\dot a^\dagger>}{\rm cm/y}$} \\
\hline\noalign{\smallskip}
%          &        &         \\
 1.5      & 5.7982 & 4.5319  \\
 2.0      & 3.8383 & 3.2801  \\
 2.5      & 2.8439 & 2.5512  \\
 3.0      & 2.2587 & 2.0873  \\
 3.5      & 1.8733 & 1.7662  \\
 4.0      & 1.6002 & 1.5307  \\
 4.5      & 1.3966 & 1.3506  \\
 5.0      &        & 1.2085  \\
 6.0      &        & 0.99830 \\
 7.0      &        & 0.85041 \\
10.0      &        & 0.58874 \\
$+\infty$ & 0      & 0       \\
\noalign{\smallskip}
\hline
\end{tabular}
\end{center}
\end{table*}
All features appearing in Fig.\,\ref{f:t108}
are included within $0\le(t_a-t)/$Gyr$\le0.25$, but dotted straight lines
define a special point, (0.198499271, 58.85857007), on the dashed curve
therein, for which the ratio of mean lunar distance to lunar diameter reads
$a/(2R_\bigcirc)=108$.   Dashed vertical straight lines mark the selected
transition age, $(t_a-t^\dagger)/$Gyr$=0.57$ (left) and the assumed EMS age,
$(t_a-t_i)/$Gyr$=4.57$ (right).   The Roche limit,
$a_{\rm R}/R_\oplus=2.898518$, is defined by a horizontal straight line.
Values inferred from paleontological data from earlier investigations
\cite{wil00} are plotted as diamonds, related to different authors when
appropriate \cite{wil00}.   Source of data (from the left to the
right) are: ER, BC (lower value preferred), WW (upper value preferred)
\cite{wil00}.

\begin{figure*}[t]               
\begin{center}                   
\includegraphics[scale=0.8]{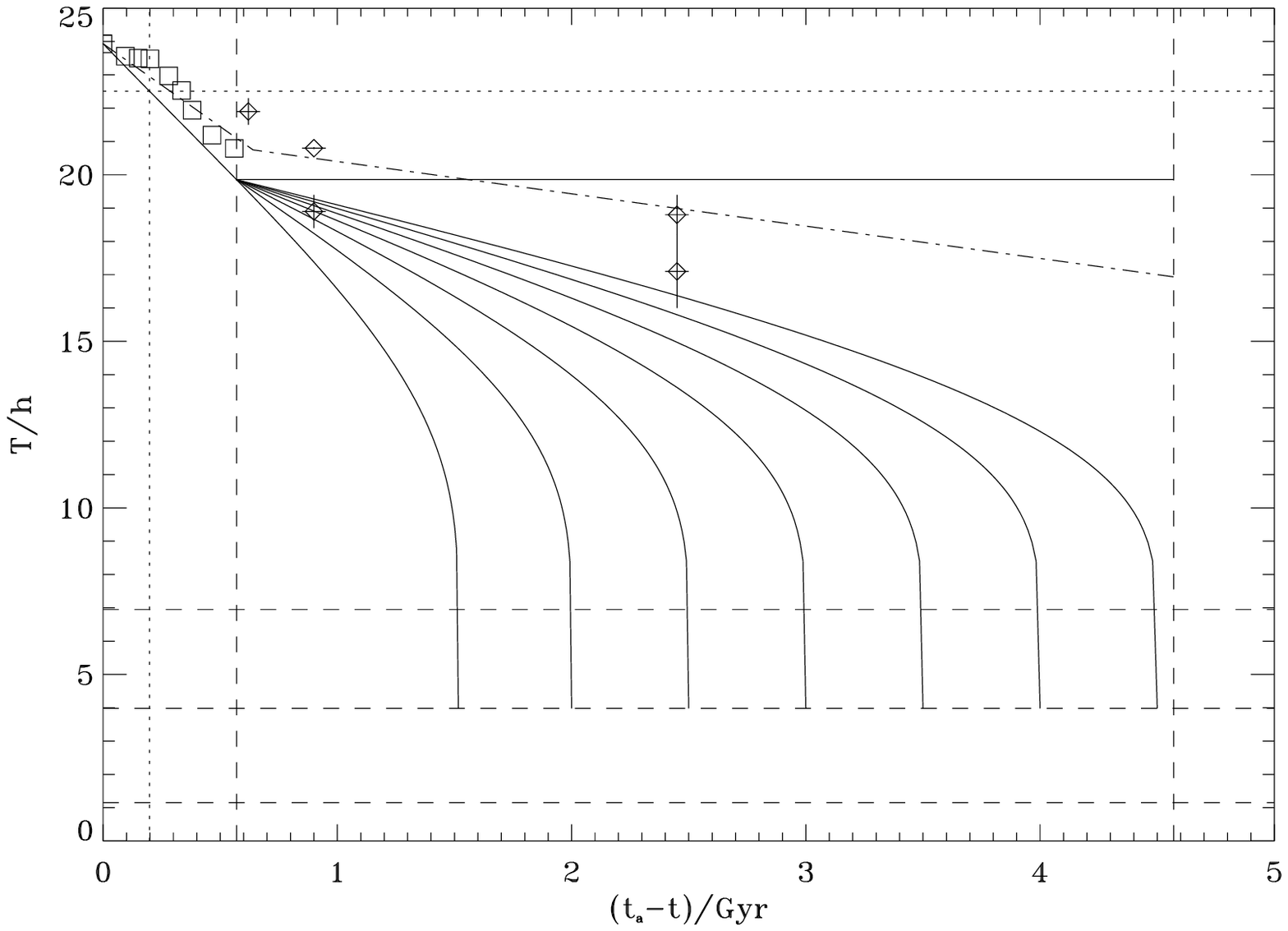}                      
\caption[ddbb]{Predicted length of day (LOD) due to tidal friction, $T/$h,
vs age (Gyr, back in time since J2000.0 date), $t_a-t$, for past 4.57 Gyr,
according to Eq.\,(\ref{eq:T}), full lines, for different merging ages (from
the left to the right), $(t_a-t^\ddagger)/$Gyr$=1.516617$, 2.0, 2.5, 3.0, 3.5,
4.0, 4.5, $+\infty$ (no lunar recession).   All features appearing in
Fig.\,\ref{f:tt108}
are included within $0\le(t_a-t)/$Gyr$\,\le0.25$, but dotted straight lines
define a special point, (0.198499271, 22.50719109), on the dashed curve
therein, for which the ratio of mean lunar distance to lunar diameter reads
$a/(2R_\bigcirc)=108$.   Dashed vertical straight lines mark the selected
transition age, $(t_a-t^\dagger)/$Gyr $=0.57$ (left) and the assumed age of
the Earth-Moon system (EMS), $(t_a-t_i)/$Gyr$ =4.57$ (right).   The Roche
limit, $T_{\rm R}/{\rm h}=6.948719$, the merging limit,
$T^\ddagger/{\rm h}=3.982278$, and the Earth Poincar\'e limit,
$T_{\rm P}/{\rm h}=1.149728$, are defined by dashed horizontal straight lines.
Dot-dashed lines are linear interpolations from paleontological data
\cite{vaa98}\cite{dea11}, extended back to $(t_a-t_i)/$Gyr$=4.57$.   A
nonlinear trend also inferred from paleontological data
\cite{vaa98}\cite{dea11} is shown as squares.   Values inferred from
accurately studied paleontological data are plotted as diamonds, related to
different authors when appropriate \cite{wil00}.   Source of data (from the
left to the right): ER, BC (lower value preferred), WW (upper value preferred)
\cite{wil00}.   See text (Introduction) for further details.}
\label{f:ut08}     
\end{center}       
\end{figure*}

Predicted LOD due to tidal friction, $T/$h, vs age, $(t_a-t)/$Gyr, according
to Eq.\,(\ref{eq:T}), is shown in Fig.\,\ref{f:ut08} as full lines
for a transition age, $(t_a-t^\dagger)/$Gyr$=0.57$, and
selected merging ages, $(t_a-t^\ddagger)$, or selected mean lunar
recession velocities just after the transition age, $<\dot a^\dagger>$, via
Eq.\,(\ref{eq:t0}), as listed in Table \ref{t:dapt}.
All features appearing in Fig.\,\ref{f:tt108}
are included within $0\le(t_a-t)/$Gyr$\le0.25$, but dotted straight lines
define a special point, (0.198499271, 22.50719109), on the dashed curve
therein, for which the ratio of mean lunar distance to lunar diameter reads
$a/(2R_\bigcirc)=108$.

Dashed vertical straight lines mark the selected
transition age, $(t_a-t^\dagger)/$Gyr$=0.57$ (left) and the assumed EMS age,
$(t_a-t_i)/$Gyr$=4.57$ (right).   Symbol captions are as in
Fig.\,\ref{f:tt108} (squares) and Fig.\,\ref{f:u108} (diamonds).   Dashed
horizontal straight lines mark the Roche limit,
$T_{\rm R}/{\rm h}=6.948719$, the merging limit,
$T^\ddagger/{\rm h}=3.982278$, and the Earth Poincar\'e limit,
$T_{\rm P}/{\rm h}=1.149728$.

Dot-dashed lines are linear interpolations from paleontological data,
$T/$h$ =23.934468-4.98(t_a-t)/$Gyr, $0\le(t_a-t)/$Gyr $\le0.64$, and
$T/$h$ =21.368068-0.97(t_a-t)/$Gyr, $0.64\le(t_a-t)/$Gyr $\le2.50$
\cite{vaa98}\cite{dea11}, extended back to $(t_a-t_i)/$Gyr$=4.57$.

\begin{figure*}[t]               
\begin{center}                   
\includegraphics[scale=0.8]{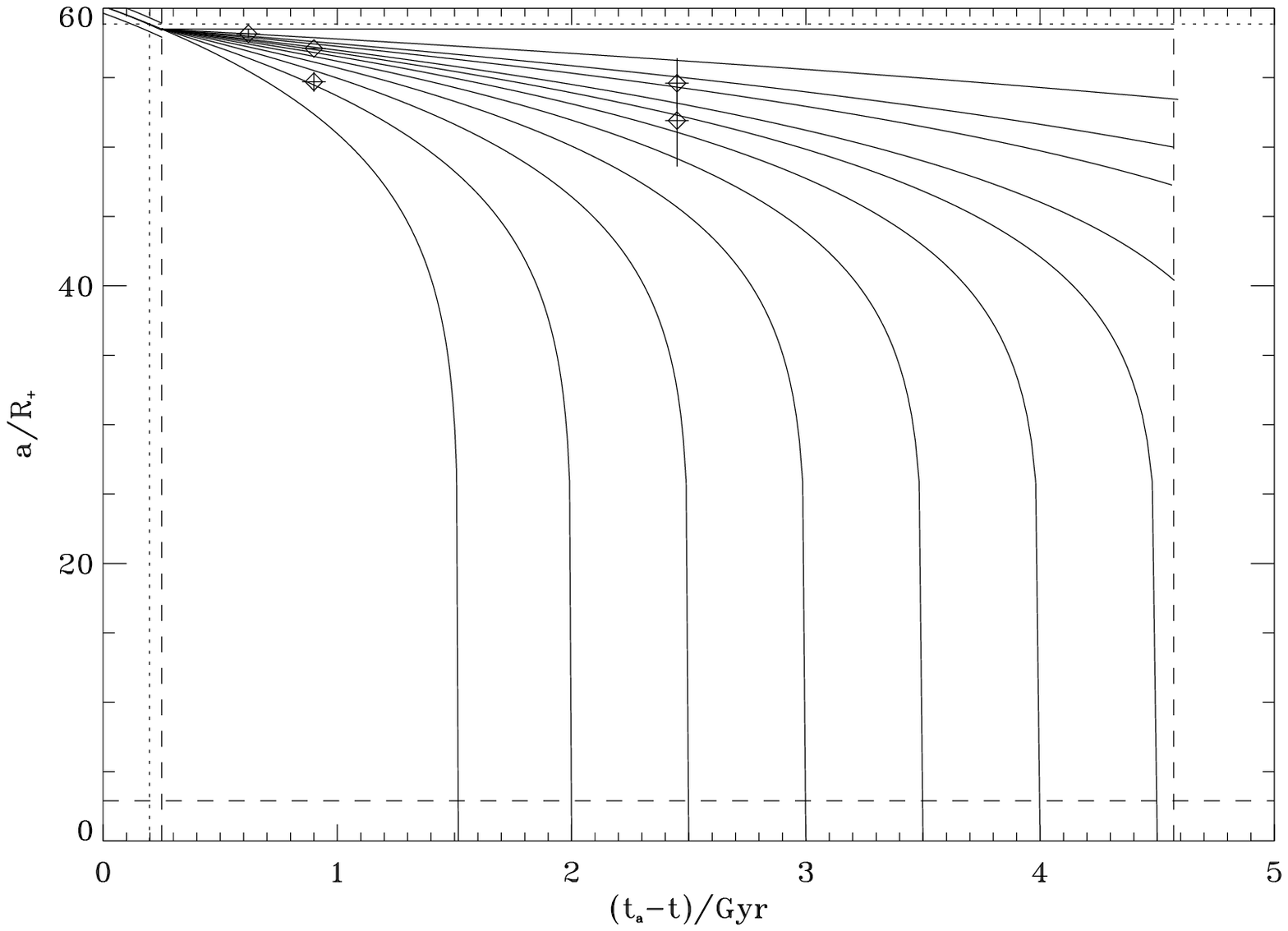}                      
\caption[ddbb]{Same as in Fig.\,\ref{f:u108} but with regard to a later
transition age, $(t_a-t^\dagger)/$Gyr $=0.25$.}
\label{f:v108}     
\end{center}       
\end{figure*}

\begin{figure*}[t]               
\begin{center}                   
\includegraphics[scale=0.8]{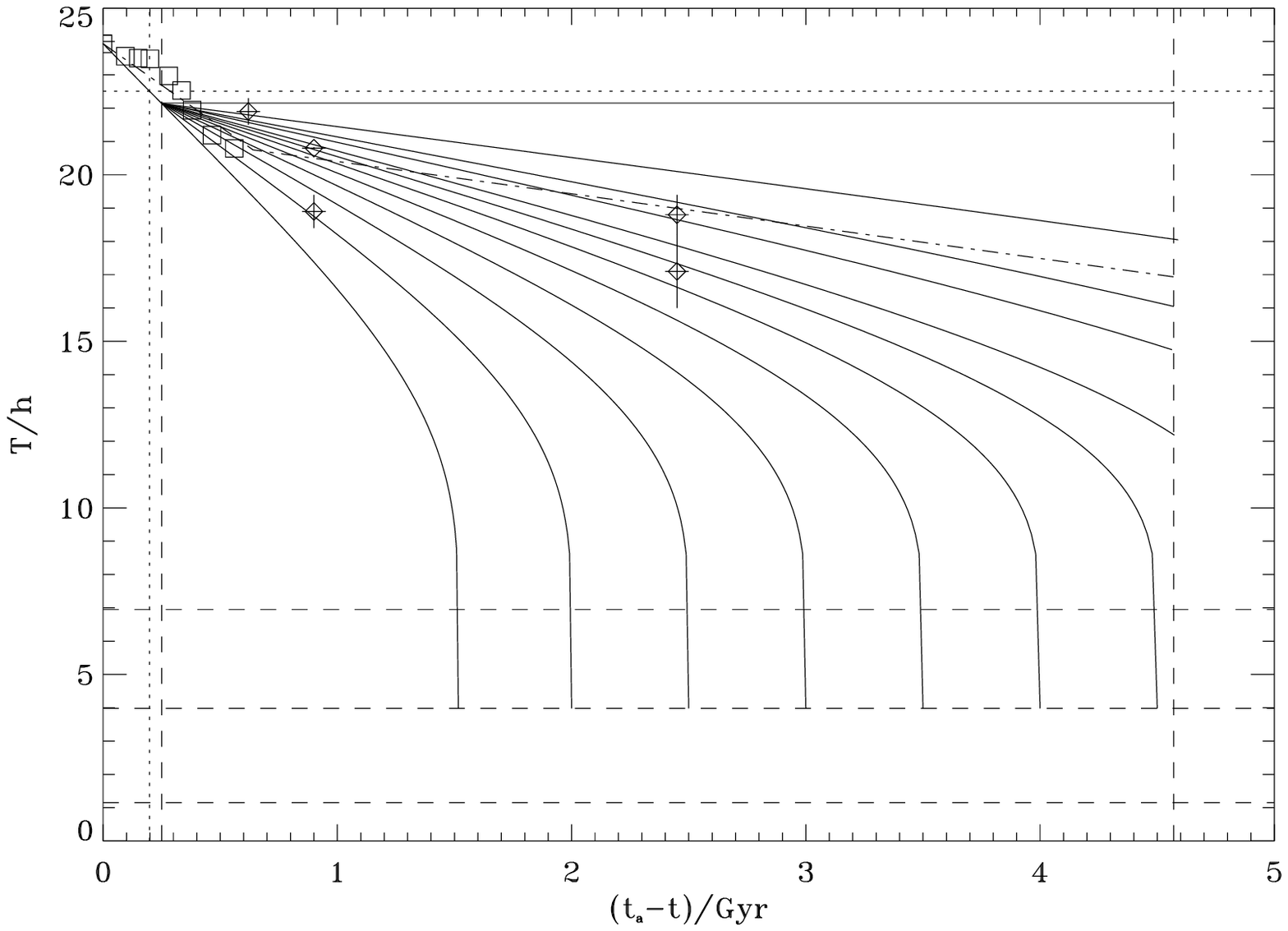}                      
\caption[ddbb]{Same as in Fig.\,\ref{f:ut08} but with regard to a later
transition age, $(t_a-t^\dagger)/$Gyr $=0.25$.}
\label{f:vt08}     
\end{center}       
\end{figure*}

An inspection of Figs.\,\ref{f:u108} and \ref{f:ut08} discloses the most
accurately inferred ER value \cite{wil00} cannot be fitted to model curves
even in the extreme case of no lunar recession after transition age (full
horizontal straight line).   Clearly the situation is reversed if the
transition age is closer to the present.   Then a later transition age,
arbitrarily chosen as $t_a-t^\dagger=0.25$ Gyr, has to be considered.   The
results are plotted in Figs.\,\ref{f:v108} and \ref{f:vt08}, where captions
are the same as in Figs.\,\ref{f:u108} and \ref{f:ut08}, respectively.

Curves related to very early merging ages,
$5\appleq(t_a-t^\ddagger)/$Gyr $\appleq10$, are found to be consistent with ER
and preferred WW value.   The related range of mean lunar recession velocity
just after the transition age, via Table \ref{t:dapt} reads
$1.21\appgeq<\dot a^\dagger>/$(cm/y) $\appgeq0.60$, about 3-6 times lower than
the current value.

Curves within a narrower range, $5\appleq(t_a-t^\ddagger)/$Gyr $\appleq7$, are
found to be
consistent with both ER, preferred WW and unpreferred BC value \cite{wil00},
or $1.2\appgeq<\dot a^\dagger>/$(cm/y) $\appgeq0.89$ via Table \ref{t:dapt},
where $<\dot a^\dagger>/$(cm/y) $\approx4.5$ just before transition.

Further restriction arises in dealing with LOD, Fig.\,\ref{f:vt08}, where
$5\appleq(t_a-t^\ddagger)/$Gyr $\appleq6$ is found following a similar
procedure, or $1.2\appgeq<\dot a^\dagger>/$(cm/y) $\appgeq0.99$, which could
be shifted downwards if processes different from solid Earth + ocean tidal
friction are effective in lowering LOD.

\section{Discussion}
\label{disc}
                                                    
%\noindent\noindent
Results from models of lunar recession need some caution, in the sense related
computer runs start at present and go back into the past.   A necessary
restrictive assumption is that minor mergers and close encounters due to
large-mass asteroids had
negligible effects on EMS evolution.   Models implying merging ages in advance
with respect to formation age, $t_a-t^\ddagger<t_a-t_i\approx4.5$\,Gyr, must
be disregarded, which is the case of continuous mean lunar recession
velocity, as shown in Figs.\,\ref{f:u108} and \ref{f:v108}.   Accordingly, key
input parameters are the merging age and the transition age, $t_a-t^\dagger$,
where mean lunar recession velocity is suddenly reduced.

Remaining input parameters are taken or inferred from a solution for
astronomical computation of insolation quantities on Earth within the range,
$-0.25\le(t_a-t)/{\rm Gyr}\le+0.25$ \cite{laa04}, to which the prediction of
the current model \cite{waz86}\cite{wil00} satisfactorily fit, as shown in
Figs.\,\ref{f:t108} and \ref{f:tt108}.   In particular, input current mean
lunar recession velocity and LOD variation read $<\dot a_a>=3.89$\,cm/y and
$\dot T_a=2.68$\,ms/cy, respectively \cite{laa04}, against
$<\dot a_a>=(3.82\mp0.07)$\,cm/y inferred from lunar laser ranging experiment
\cite{dia94} and $\dot T_a=(2.50\mp0.01)$\,ms/cy determined from solid Earth +
ocean tidal friction \cite{mal09}.

Predicting LOD implies further problems, as lunar recession occurs only via
tidal friction while LOD depends, in addition, on atmospheric tides and
nontidal processes e.g., \cite{vaa98}\cite{laa04}\cite{dea11}.   The current
model works in presence of tidal friction only, which implies a LOD lower
limit, keeping in mind the above mentioned additional effects act in the
opposite direction i.e. increasing Earth spin.   On the other hand, the model
can be extended to the special case where the contribution from atmospheric
tides and nontidal processes is time independent i.e. related LOD variation
maintains the current value,
$\Delta\dot T_a=(-0.10-0.55)$\,ms/cy $=-0.65$\,ms/cy \cite{mal09}.

A comparison between model predictions and paleontological data inferred from
fossil corals and molluscs e.g.,
\cite{scr78}\cite{lam80}\cite{waz86}\cite{soa96b}\cite{sac98}\cite{wil89a}\cite{wil00}
can be made in a twofold manner, according if lunar recession or LOD are
involved.

With regard to lunar recession, empirical values are inferred from the
following formations: WW \cite{waz86}\cite{wil89c}\cite{wil90}, BC
\cite{soa96b}\cite{sac98}, ER
\cite{wil89a}\cite{wil89b}\cite{wil89c}\cite{wil90}\cite{wil91}\cite{wil94}\cite{wil97},
where a collection appears in a later investigation \cite{wil00}.   For a
transition age fixed at the late Precambrian, $t_a-t^\dagger=0.57$\,Gyr, even
if no lunar recession occurs at earlier ages, predicted EMD cannot fit to ER
and WW data, while cases where the merging age is not in advance with respect
to the formation age, $t_a-t^\ddagger\ge t_a-t_i\approx4.5$\,Gyr, are
marginally consistent with BC data, as shown in Fig.\,\ref{f:u108}.

Conversely, a transition age selected on the late Permian,
$t_a-t^\dagger=0.25$\,Gyr, yields acceptable cases,
$5\le(t_a-t^\ddagger)/$\,Gyr $\le7$, where predicted EMD fits to ER, BC
(unpreferred value) and WW (preferred value) data, as shown in
Fig.\,\ref{f:v108}.   In addition, acceptable cases cannot fit to BC
(preferred value) data, which implies at least one model assumption has to be
released, provided data under discussion are correct, namely mean lunar
recession velocity underwent several (instead of a single) discontinuities in
the past.

The above mentioned discrepancy could be lowered by the occurrence of
short-period $(\Delta t\approx10^{-3}$\,Gyr) EMD fluctuations
$(\mp\Delta a\approx0.5R_\oplus)$, which were established back to 0.25 Gyr via
astronomical computations \cite{laa04}.   If fluctuations of the kind
considered took place back to 2.50 Gyr, it cannot be excluded EMD inferred
from ER and WW data relate to a fluctuation maximum i.e. in excess of about
$0.5 R_\oplus$ with respect to related mean EMD, and EMD inferred from BC data
relate to a fluctuation minimum i.e. in defect of about $0.5 R_\oplus$ with
respect to related mean EMD.   Accordingly, the preferred EMD inferred from
BC data \cite{wil00} should be incremented by about $1R_\oplus$, passing from
$(54.7\mp0.7)R_\oplus$ to $(55.7\mp0.7)R_\oplus$, which could be consistent
with the unpreferred EMD inferred from BC data, equal to $57.1R_\oplus$ where
the error is not reported \cite{wil00}.

Earlier applications of the current model yield similar results e.g.,
\cite{waz86}\cite{wil00}, regardless of (slightly different) values of input
parameters.   The current choice, listed in Table \ref{t:inpu}, makes a set of
homogeneous data from a single source \cite{laa04}, which ensures consistency
with astronomical computations, as shown in Figs.\,\ref{f:t108} and
\ref{f:tt108}.

With regard to LOD, in addition to ER, BC, and WW data, different linear
interpolations are available for Proterozoic and Phanerozoic e.g.,
\cite{vaa98}\cite{dea11}, as:
\begin{leftsubeqnarray}
\slabel{eq:lipa}
&& T=T_a-4.98(t_a-t)~~;\qquad0\le t_a-t\le t_a-t^\ast~~; \\
\slabel{eq:lipb}
&& T=T_a^\prime-0.97(t_a-t)~~;\qquad t_a-t^\ast\le t_a-t\le t_a-t_i
\label{seq:lip}
\end{leftsubeqnarray}
where $T_a=23.934468$\,h is the current (2000.0 JC date) LOD \cite{laa04},
$T$ is the LOD at the age, $t_a-t$, and $t_a-t^\ast=0.64$\,Gyr.

The intercept, $T_a^\prime$, can be determined keeping in mind
Eqs.\,(\ref{eq:lipa}) and (\ref{eq:lipb}) both hold at the intersection point
of related straight lines.   The result is:
\begin{lefteqnarray}
\label{eq:Tap}
&& T_a^\prime=T_a-(4.98-0.97)(t_a-t^\ast)=21.368068\,{\rm h}~~;
\end{lefteqnarray}
and, in addition, $T^\ast=T_a-4.98(t_a-t^\ast)=20.747268\,{\rm h}$ via
Eq.\,(\ref{seq:lip}).

Predicted LOD agrees to an acceptable extent with its counterpart inferred
from astronomical computations \cite{laa04}, with a maximum discrepancy of
12\,m at $t_a-t=0.25$\,Gyr.   The contrary holds with respect to inferred
(from paleontological data) LOD, Eq.\,(\ref{eq:lipa}), where the maximum
discrepancy amounts to 30\,m at the same age, which is worsened considering a
nonlinear trend, as shown in Fig.\,\ref{f:tt108}.

This is why predicted LOD
is affected only by tidal friction in models under discussion
\cite{waz86}\cite{wil00}\cite{laa04}, while inferred (from paleontological
data) LOD also includes atmospheric tides and nontidal processes such as
mantle convection and iron macrodiffusion, which act in opposite sense with
respect to tidal friction e.g.,
\cite{vaa98}\cite{laa04}\cite{mal09}\cite{dea11}.

More specifically, the contribution from total solid Earth + ocean to current
LOD variation is evaluated as $(2.50\mp0.01)$\,ms/cy, and the contribution
from atmospheric tides and nontidal processes as $(-0.10\mp0.01)$\,ms/cy and
$(-0.55\mp0.01)$\,ms/cy, respectively \cite{mal09}.   Then the global
contribution to current LOD variation amounts to $(1.85\mp0.01)$\,ms/cy.
Accordingly, the input current LOD variation, $\dot T_a=2.68$\,ms/cy, relates
to the contribution from total solid Earth + ocean, while empirical LOD
variation inferred from linear interpolation,
$\dot T_a=4.98$\,h/Gyr $\approx1.79$\,ms/cy, relates to the global
contribution.

Under the restrictive assumption of time independent contribution from
atmospheric tides and nontidal processes, $\Delta T/(t_a-t)=$ const, using the
above mentioned values yields $\Delta T/(t_a-t)=0.65$\,ms/cy, or:
\begin{lefteqnarray}
\label{eq:DT}
&& \frac{\Delta T}{\rm h}=\frac{65}{36}\frac{t_a-t}{\rm Gyr}~~;
\end{lefteqnarray}
where $T+\Delta T$ is the LOD at the age, $t_a-t$, due to total solid Earth +
ocean tidal friction $(T)$ with the addition of atmospheric tides and
nontidal processes $(\Delta T)$.   Related explicit form via 
Eqs.\,(\ref{eq:T}) and (\ref{eq:DT}) reads:
\begin{lefteqnarray}
\label{eq:TDT}
&& \frac{T+\Delta T}{T_a}=C_1\left\{C_3-\left(\frac{a}{a_a}\right)^{1/2}\left[
1+\frac{C_2}{13}\left(\frac{a}{a_a}\right)^6\right]\right\}^{-1}+\frac{65}{36}
\frac{t_a-t}{T_a}\frac{\rm h}{\rm Gyr}~~;\quad
\end{lefteqnarray}
which, in the special case of merging age, $a=0$, $T=T^\ddagger$, reduces to:
\begin{lefteqnarray}
\label{eq:TDT0}
&& T^\ddagger+\Delta T^\ddagger=\frac{C_1}{C_3}T_a+\frac{65}{36}
(t_a-t^\ddagger)=3.982659\,{\rm h}+\frac{65}{36}(t_a-t^\ddagger)
\frac{\rm h}{\rm Gyr}~~;
\end{lefteqnarray}
that is a straight line in the variables, $T^\ddagger$ and $t_a-t^\ddagger$.

Turning to the general case, the curve described by  
Eq.\,(\ref{eq:TDT}) is plotted in Fig.\,\ref{f:tt108} as a dotted line, which
is satisfactorily close to the interpolation line (dot-dashed) inferred from
paleontological data, Eq.\,(\ref{eq:lipa}).   Fitting to an inferred nonlinear
trend \cite{vaa98}\cite{dea11}, plotted in Fig.\,\ref{f:tt108} as squares,
would need several (instead of a single) discontinuities in LOD variation,
taking due account of, say, Pangaea formation and deformation e.g.,
\cite{vaa98}\cite{dea11}.

An extension of inferred (from paleontological data) interpolation line to
Archean and, fictiously, to EMS formation age, could be illustrative for
comparison with predicted LOD, as shown in Figs.\,\ref{f:ut08} and
\ref{f:vt08}.   To this aim, a choice must be done of what is the better
empirical evidence among ER, BC, WW data \cite{wil00} on one hand, and
inferred linear trend considering the above mentioned data as representative
as the remaining ones \cite{dea11} on the other hand.   In fact, inferred
linear interpolation lies below ER value and more or less in the middle of
preferred and unpreferred BC and WW values.   Both alternatives shall be
considered in the following.

If the better empirical evidence of LOD relates to ER, BC and WW data, current
models yield acceptable fits with regard to (i) late transition ages, and
(ii) unpreferred BC value, as shown in Fig.\,\ref{f:vt08} for curves related
to merging ages, $5\le(t_a-t^\ddagger)/$ Gyr $\le6$.

If the better empirical evidence of LOD relates to inferred linear
interpolations, current models provide acceptable fits for late transition
ages, as shown in Fig.\,\ref{f:vt08} for curves related to merging ages,
$(t_a-t^\ddagger)/$Gyr $\approx6$.

In addition, errors on age of ER, BC and WW formation are not
available e.g., \cite{wil00} and for this reason an uncertainty of
$\mp0.05$\,Gyr has arbitrarily been assumed in plotting related values, even
if larger uncertainties are probably expected.

\begin{figure*}[t]               
\begin{center}                   
\includegraphics[scale=0.8]{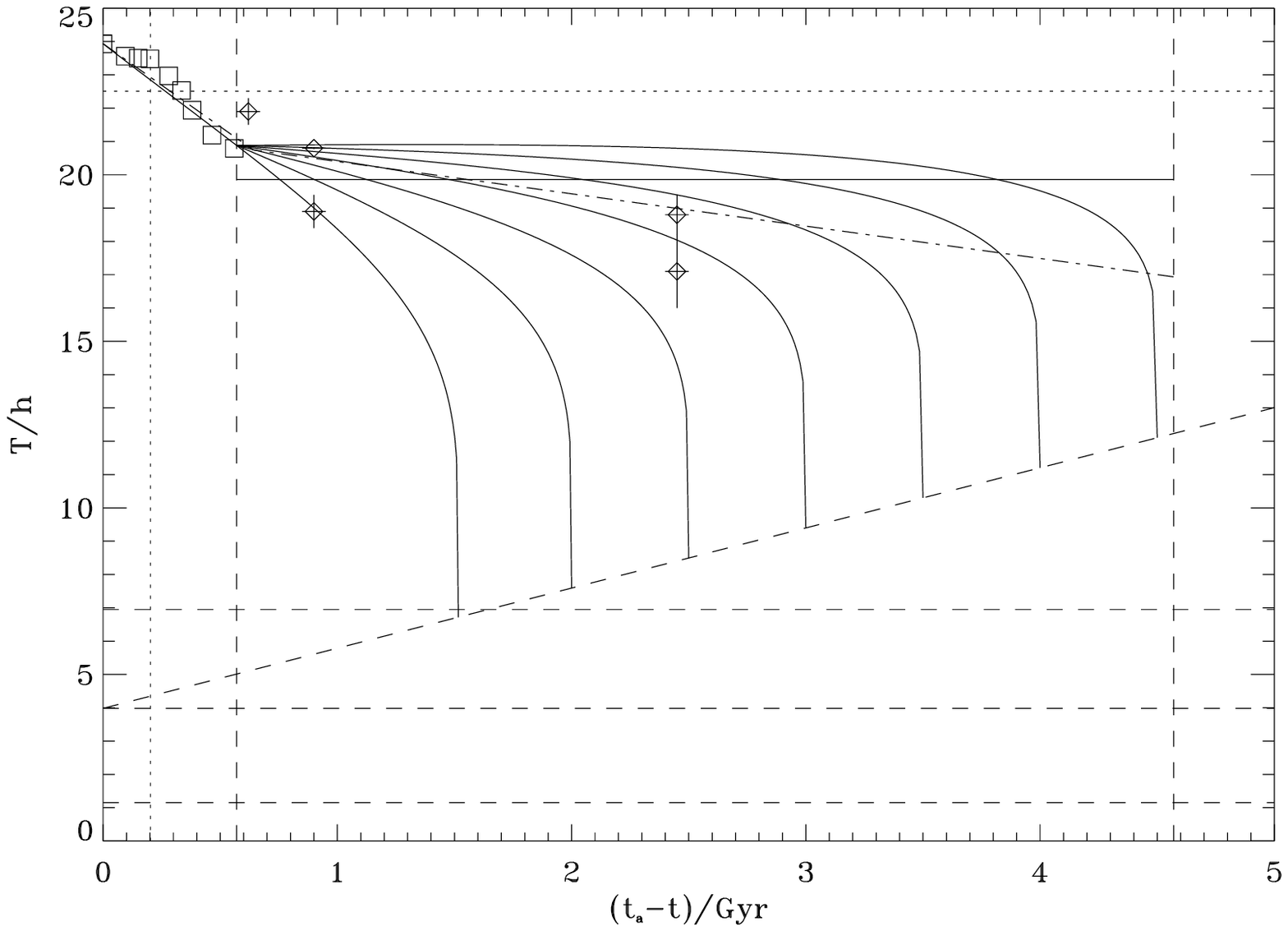}                      
\caption[ddbb]{Same as in Fig.\,\ref{f:ut08} but including, in addition to
tidal friction, the effect of atmospheric tides and nontidal processes via an
assumed time independent contribution yielding an additional term,
$\Delta T/$h $=(65/36)(t_a-t)/$Gyr.
The locus of merging configurations $(a=0)$ is shown as an inclined dashed
straight line.   See text for further details.}
\label{f:utt8}     
\end{center}       
\end{figure*}

\begin{figure*}[t]               
\begin{center}                   
\includegraphics[scale=0.8]{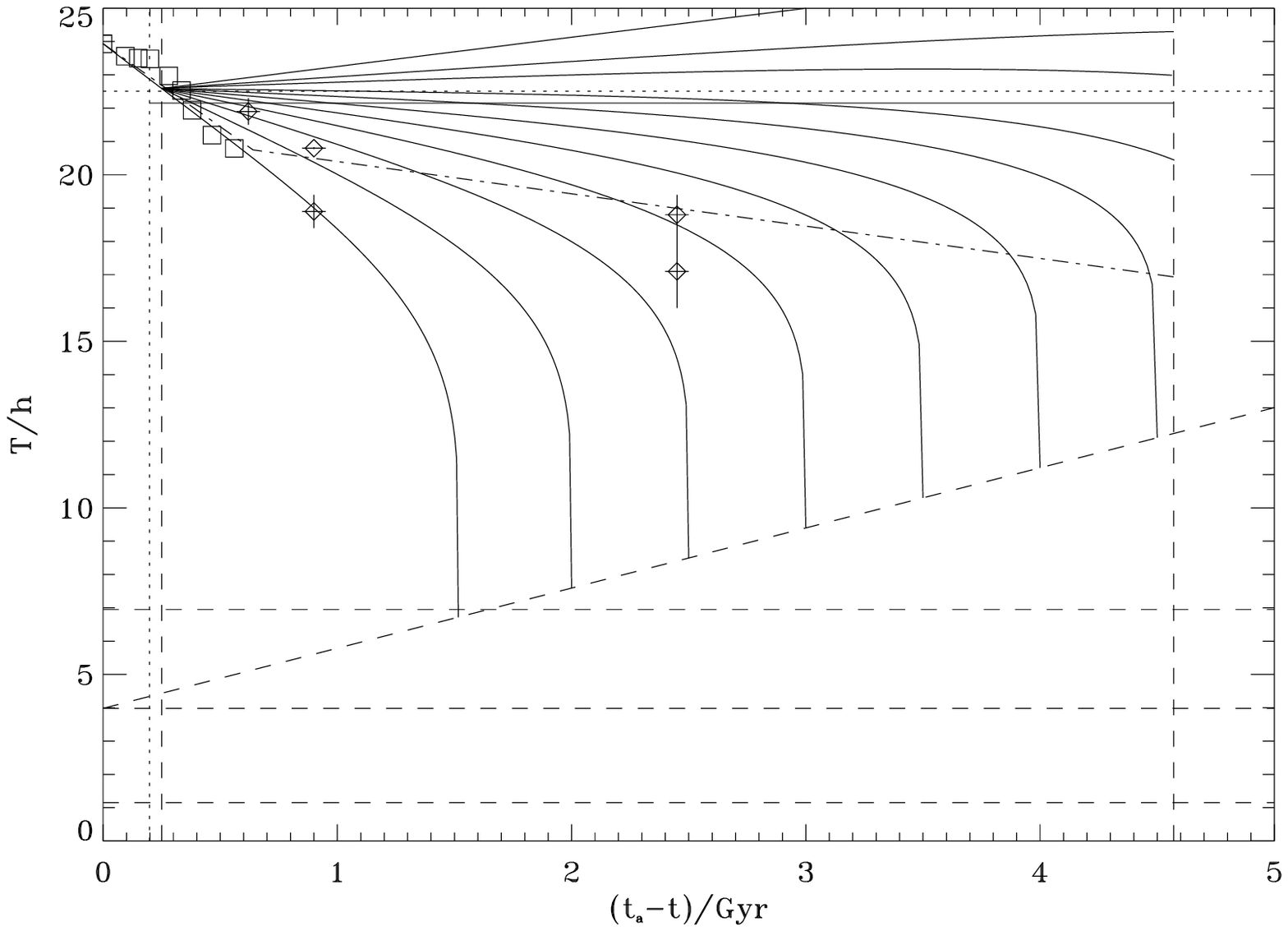}                      
\caption[ddbb]{Same as in Fig.\,\ref{f:vt08} but including, in addition to
tidal friction, the effect of atmospheric tides and nontidal processes via an
assumed time independent contribution yielding an additional term,
$\Delta T/$h $=(65/36)(t_a-t)/$Gyr.
The locus of merging configurations $(a=0)$ is shown as an inclined dashed
straight line.   See text for further details.}
\label{f:vtt8}     
\end{center}       
\end{figure*}

Keeping into due account
the (time independent) effect of atmospheric tides and nontidal processes, the
situation is worsened as shown in Figs.\,\ref{f:utt8} and \ref{f:vtt8}, in the
sense the merging age would be in advance with respect to the formation age.
To avoid this inconsistency, the above mentioned effect should increase going
back in time.

In conclusion, predicted LOD satisfactorily fits to both ER, BC, WW data and
inferred linear interpolations with regard to late transition ages,
$t_a-t^\dagger\approx0.25$\,Gyr, and merging ages,
$t_a-t^\ddagger\approx6$\,Gyr, leaving aside the effect of atmospheric tides
and nontidal processes.   Suitable choices of the above mentioned input
parameters could better reproduce empirical values, but the main result is
current models are consistent with the data in spite of restrictive
assumptions.

On the other hand, a linear LOD-age relation is a zeroth-order approximation,
as shown in Figs.\,\ref{f:tt108}, \ref{f:ut08}, \ref{f:vt08}, \ref{f:utt8},
\ref{f:vtt8}, by comparison with data from earlier investigations
\cite{vaa98}\cite{dea11},
plotted as squares.   A nonlinear trend could be owing to e.g., occurrence and
persistence of a supercontinent associated to a deep superocean
\cite{vaa98}\cite{dea11}.   Even in this context, the current model could fit
to the data if several (instead of a single) transition ages are considered,
where mean lunar recession velocity suddenly drops to a lower value.   To
this aim, the effect (impacts or close encounters) of large-mass asteroids 
should be analysed in detail.

\section{Conclusion}
\label{conc}

%\noindent\noindent
The evolution of Earth-Moon system (EMS) relates to several different
disciplines, such as Astronomy, Geodesy, Geophysics, Oceanology, Geology,
Paleontology, where comparison between theoretical predictions and results
from data collections could imply constraints within all the involved fields.
In this view, a classical model \cite{waz86}\cite{deu90}\cite{wil00} has been
reviewed and updated to many extents.

First, a homogeneous set of input parameters has been taken from earlier
astronomical computations on insolation quantities on Earth spanning from
$-$0.25\,Gyr to +0.25\,Gyr \cite{laa04}.

Second, predicted Earth-Moon distance (EMD) for assigned transition age,
$t_a-t^\dagger$, and merging age, $t_a-t^\ddagger$, has been considered
keeping in mind acceptable cases necessarily imply merging age in advance
with respect to formation age, $t_a-t_i$.

Third, the presence of short-period $(\Delta t\approx10^{-3}$\,Gyr) EMD
fluctuations $(\Delta a\approx\mp0.5R_\oplus)$ from astronomical
computations \cite{laa04}, if typical of the whole evolution, implies
uncertainty bands around $a(t)$ curves of thickness about $\mp0.5R_\oplus$
and values inferred from paleontological data affected by an additional error,
$\Delta a\approx\mp0.5R_\oplus$, if related to mean (over a typical
fluctuation period) EMD.

The main results of the present paper can be summarized as follows.
\begin{description}
\item[(1)~] 
Predicted EMD and length of day (LOD) slightly overstimate their counterparts
from astronomical computation of insolation quantities on Earth back to
0.25\,Gyr \cite{laa04}.
\item[(2)~] 
Predicted LOD, when a time independent effect via atmospheric tides and
nontidal processes is considered \cite{mal09}, slightly understimates a linear
interpolation from paleontological data related to Phanerozoic era
\cite{vaa98}\cite{dea11}.
\item[(3)~]
For assigned transition age and merging age, predicted EMD and LOD cannot fit
to their counterparts inferred from ER, BC and WW data, unless unpreferred
instead of preferred values \cite{wil00} are considered for BC data.   Related
difference in EMD appears comparable to amplitude in short-period fluctuations
from astronomical computations during the late Phanerozoic \cite{laa04}.
Accordingly, a transition age, $t_a-t^\dagger=0.25$\,Gyr and a merging age,
$t_a-t^\ddagger=6$\,Gyr, yield an acceptable fit.   The situation is reversed
if a time independent effect via atmospheric tides and nontidal processes is
considered, in the sense that fitting curves relate to merging age in  advance
with respect to formation age.
\item[(4)~]
For assigned transition age, $t_a-t^\dagger=0.25$\,Gyr and merging age,
$t_a-t^\ddagger=6$\,Gyr, predicted LOD is consistent with inferred linear
interpolation to paleontological data \cite{vaa98}\cite{dea11}. The  situation
is reversed  if a time independent effect via atmospheric tides and nontidal
processes is considered, in the sense that fitting curves relate to merging
age in  advance with respect to formation age.
\end{description}
Even if LOD variation was not linear in time e.g., \cite{vaa98}\cite{dea11},
the current model remains valid as a zero-th order approximation.   Further
refinement would be desirable in dealing with paleontological data, in
connection with error evaluation.

\section*{Acknowledgements}

Thanks are due to S. Lambert for fruitful e-mail correspondance.

\end{document}